\documentclass[%
showkeys,
preprint,
amsmath,amssymb,
aps,
]{revtex4-1}
\usepackage{graphics}
\usepackage{epsfig}
\usepackage{epstopdf}
\usepackage{amsmath}
\usepackage{array}
\usepackage{subfigure}
\usepackage[colorlinks=true,linkcolor=blue,citecolor=blue,      urlcolor=blue]{hyperref}
\begin{document}
    \title{First Law of Thermodynamics and Emergence of Cosmic Space in a Non-Flat Universe}
	\author{Hareesh T}
	\email{hareesht23@gmail.com}
	\author{P. B. Krishna}
	\email{krishnapb@cusat.ac.in}
	\author{Titus K. Mathew.}
	\email{titus@cusat.ac.in}
	\affiliation{%
		Department of Physics, Cochin University of Science and Technology, Kochi, Kerala 682022, India
		}%
\begin{center}
\begin{abstract}
	The emergence of cosmic space as cosmic time progresses is an exciting idea advanced by Padmanabhan to explain the accelerated expansion of the universe. The generalization of Padmanabhan's conjecture to the non-flat universe has resulted in scepticism about the choice of volume such that the law of emergence can not be appropriately formulated if one uses proper invariant volume. The deep connection between the first law of thermodynamics and the law of emergence \cite{mahith}, motivate us to explore the status of the first law in a non-flat universe when one uses proper invariant volume. We have shown that the first law of thermodynamics, $dE = TdS +WdV$ cannot be formulated properly for a non-flat universe using proper invariant volume. We have also investigated the status of the first law of the form $-dE = TdS$ in a non-flat universe. We have shown that the energy change dE within the horizon and the outward energy flux are not equivalent to each other in a non-flat universe when we use the proper invariant volume. We have further shown that the consistency between the above two forms of the first law claimed in Ref. \cite{caiakb} will hold only with the use of the areal volume of the horizon. Thus, a consistent formulation of the above two forms of the first law of thermodynamics demands the use of areal volume.

\end{abstract}

\maketitle

\end{center}
	\section{Introduction}
The connections between gravity and thermodynamics have emerged as an important area of research since the development of black-hole thermodynamics \cite{Bardeen,tem,entropy}. These captivating results are not mere coincidences but have lead to more profound insights into the quantum nature of gravity. Jacobson was the first to discover the direct connections between gravity and thermodynamics. He derived Einstein's field equations from the Clausius relation, $\delta Q=TdS$ at the horizon together with the equivalence principle, where $\delta Q$ and $T$ are the energy flux and Unruh temperature perceived by an accelerated observer within the horizon, respectively \cite{grt}.
Further investigations on the thermodynamic perspective of the gravitational field equations have brought more light into the connections between gravity and thermodynamics \cite{Physical_Interpretation_Gravitational_Equations,lovelockfrstlaw,Paddy_thrmodynamics_gravity,thermodynamics_gravity,paddyreview1}. These close resemblances between gravity and thermodynamics strongly point to a microscopic structure for gravity, and abundant pieces of evidence for gravity to be an emergent phenomenon \cite{paddyreview,Paddy_emergnt_gravity,Paddy_emergnt_gravity1,emergnt_gravity,emergnt_gravity1,emergnt_gravity2}. On this background, Verlinde treated gravity as an entropic force and successfully derived Newton's gravitation law  \cite{verlinde}. Similarly, Padmanabhan also derived the law of gravitation using equipartition law of energy and the thermodynamic relation $S=\frac{E}{2T}$, where S is the
thermodynamic entropy, T the temperature of the horizon and E is the gravitational mass \cite{thanugrt2,Paddyentropy}.

In these earlier treatments, the space-time is assumed to be pre-existing, and the field equations arose as consistency conditions obeyed by the background space-time. The question now to address is how reasonable it is to assume the space-time itself is an emergent structure. Padmanabhan tackled the conceptual complexities raised by this question by considering a geodesic observer in cosmology. To this observer, the expansion of the universe would be equivalent to the emergence of cosmic space as cosmic time progresses. Coupling this idea with a specific version of the holographic principle, he further postulated that the expansion of our universe is driven towards the holographic equilibrium\cite{thanu}. Following this idea, the further perspectives of emergent paradigm were explored on \cite{emergent_paradigm_invsetigation1,emergent_paradigm_invsetigation2,emergent_paradigm_invsetigation3,krishnapb1,krishnapb,krishnapb2} and for recent investigations on this novel idea refer \cite{Recent_investigation1,Recent_investigation2,Recent_investigation3,Bassari_model}. Based on the holographic equipartition, Padmanabhan proposed an expansion law for flat (3+1) dimensional universe by showing it leads to the Friedmann equations \cite{thanu}. Cai extended the expansion law in the Einstein gravity to Gauss-Bonnet and the more general Lovelock gravity models by appropriately modifying the surface degrees of freedom and the effective volume of the horizon.
\cite{rgcaiflathighD}. With a minor correction to the expansion law, Sheykhi was successful in producing the Friedmann equations of a universe with any curvature \cite{sheykhi}. 
Even though the modified expansion law due to Sheykhi was successful in deriving the Friedmann equations, the use of areal volume even for the non-flat universe was criticized. Because it leads to the conclusion that the time evolution of the flat universe generates non-flat Friedmann equations \cite{modthanu}. Moreover, it is reasonable to assert that the geometrical quantities are metric dependent, and hence, the proper invariant volume should be favoured instead of the areal volume.  Motivated by these objections, Eune and Kim employed the proper invariant volume to formulate the expansion law in a non-flat universe. They reproduced the Friedmann equations with invariant volume but had to redefine the Planck length as a function of cosmic time. Moreover,the redefined Planck length, $(l_p^2)^{\text{eff}}$ is divergent for radiation and matter-dominated era for $k=-1$, while it approaches $(l_p^2)^{\text{eff}}_{k=1} =2l_P^2$ for $k=1$ and these results are not compatible with the standard observations \cite{modthanu}.

Recently, the authors of \cite{mahith} have derived the expansion law proposed in \cite{sheykhi} from the unified first law of thermodynamics ($dE=TdS+WdV$) in (n+1) dimensional Einstein, Gauss-Bonnet and more general Lovelock gravity. Similarly, the authors of \cite{expasnion_law_fom_frst_law} have also derived the expansion law in \cite{sheykhi} from a specific version of the first law of thermodynamics ($-dE=TdS$) in (3+1) Einstein gravity. In these works, the areal volume was considered as the appropriate volume to derive the expansion law.
Thus the results presented in \cite{mahith} and \cite{expasnion_law_fom_frst_law} furnish the substantial shreds of evidence to consider the first law of thermodynamics as the backbone of expansion law. 
In this context, the previously noted challenges faced by the expansion law proposed with invariant volume motivate us to suspect the feasibility of formulating the first law of thermodynamics using the invariant volume. In this paper, we address the uncertainty posed by the choice of volume to describe the expansion law from a thermodynamic point of view using the first law of thermodynamics.

This paper is structured as follows. In section \ref{2} after brief reviews of the expansion laws introduced in \cite{sheykhi} and \cite{modthanu} we bring out the difficulties in formulating the expansion law using invariant volume without altering Planck length. In section \ref{2.1}, we discuss the status of the unified first law of thermodynamics if one uses invariant volume. In section \ref{3}, we employ the first law of thermodynamics of the form $dE=-TdS$ to derive the expansion law presented in \cite{sheykhi} and analyze how the choice of volume will affect the formulation of the first law.  In the last section, we summarise the results and present our conclusions.  We use the natural units $c=\hslash=k_b=1$ for simplicity.

\section{Feasibility of Expansion Law in a Non-Flat Universe}\label{2}
We assume the universe to be homogeneous and isotropic, described by the line element\cite{caiakb}
\begin{equation}\label{metric}
ds^2=h_{ab} dx^a dx^b + r_A^2d\Omega_{2},
\end{equation}
where $r_A=ra(t)$ and $ x^0=t,x^1=r$, $d\Omega_{2}$ denotes the line element of the 2-dimensional unit sphere and the
two dimensional metric $h_{ab}=$  diag($1,\frac{a^2}{1-kr^2})$. Here, the spatial curvature constant k=0 for a flat universe and k = $\pm1$ corresponds to a
closed and open universe respectively. In FRW cosmology, there exists a marginally
trapped surface with vanishing expansion identified by the relation $h_{ab}\partial^ar\partial^br=0$ and named
as the apparent horizon \cite{Caifrstlaw}. The thermodynamic aspects of the apparent horizon are well scrutinised and
established, leading to consider it as a suitable thermodynamic boundary of our universe \cite{Gong,frst_law_apparent_horizon,Cai_thermodynamics_apparent_horizon,Sheykhi_thermodynamics_apparent_horizon,Sheykhi_thermodynamics_apparent_horizon1}.
A straightforward calculation yields the radius of the dynamic apparent horizon as
\cite{Hayward_apparnt_horizon}
\begin{equation}\label{radius}
r_A^2=\frac{1}{H^2+\frac{k}{a^2}},
\end{equation}
where $H=\dot{a}/a$ is the Hubble parameter.
We follow the assumptions in \cite{sheykhi} and \cite{modthanu} to define the number of degrees of freedom associated with the apparent horizon in (3+1) Einstein gravity as
\begin{align}\label{degrees_of_freedom}
N_{sur}& = \frac{4\pi r_A^2}{l_p^2},\text{\quad  and} & N_{bulk}&=-4\pi r_A(\rho+3p)V,
\end{align}
where $N_{sur}$ is the number of degrees of freedom on the horizon, $N_{bulk}$ is the number of degrees of freedom in bulk and V is the volume of space inside the apparent horizon.
The novel idea advanced by Padmanabhan is that the expansion of the universe which can be considered as equivalent to the emergence of space is driven towards holographic equipartition. Mathematically this concept can be realised for the flat universe as  \cite{thanu}
\begin{equation}\label{expasnion_law_paddy}
\frac{dV}{dt}=l^2_p(N_{sur}-N_{bulk})= l^2_p\Delta N.
\end{equation}
More generally one should expect  $dV/dt$ to be a function of $ \Delta N$, which vanishes as  $\Delta N \rightarrow 0$. Then the Eq. \eqref{expasnion_law_paddy} can be interpreted as the Taylor series expansion of the original function truncated at first order \cite{thanu} (For further investigations on this line of thought refer \cite{Genralised_Expasnion_law,Genralised_Expasnion_law_modified}). Sheykhi extended this principle to non-flat universe and advocated a more general expansion law given by
\begin{equation}\label{expasnion_law_sheykhi}
\frac{dV}{dt}=l^2_p\frac{r_A}{H^{-1}}(N_{sur}-N_{bulk}).
\end{equation}
The proposed law \eqref{expasnion_law_sheykhi}  reduces to Eq.\eqref{expasnion_law_paddy} in flat space and was able to reproduce the Friedmann equations with any spatial curvature. 
Although the expansion law \eqref{expasnion_law_sheykhi} is formulated for a universe with any spatial curvature, the volume used for a non-flat universe was areal volume which resembles the volume of a sphere in Euclidean space. Furthermore, it should be noted that the areal volume can only account for the volume of space inside the apparent horizon when $k=0$. In general,  the proper invariant volume within the apparent horizon is given as 
\begin{equation}
V_k=4\pi a^3\int_{0}^{\frac{r_A}{a}}dr\frac{r^2}{\sqrt{1-kr^2}}.
\end{equation}
Eune and Kim revised the
expansion law \eqref{expasnion_law_paddy} with a  proportionality function by employing the invariant volume and derived the Friedmann equations  \cite{modthanu}. The corresponding expansion law is given by
\begin{equation}\label{ Myungseok_expasnion_law}
\frac{dV_k}{dt}=l_p^2f_k(t)\big(N_{sur}-\epsilon N_{bulk}\big),
\end{equation}
where function $f_k(t)$ is defined to be
\begin{equation}\label{fkt}
f_k(t)=\frac{\bar{V_k}[\dot{r_A}H^{-1}/r_A+(r_A/H^{-1})(H^{-1}/r_A-V_k/\bar{V_k})]}{V_k(\dot{r_A}H^{-1}/r_A+V_k/\bar{V_k}-1)},
\end{equation}
where $\bar{V_k}=\frac{4\pi r_A^3}{3}$ is the areal volume.
To maintain the original form of Eq.\eqref{expasnion_law_paddy}, the proportional function along with the Planck length in Eq.\eqref{ Myungseok_expasnion_law} was interpreted as $l_p^2f_k(t)=(l_p^2)^{\text{eff}}$, where $(l_p^2)^{\text{eff}}$ is the effective Planck length \cite{modthanu}. But, altering the fundamental constant does not seem reasonable and has attracted much criticism for it \cite{modthanu_critic} 

Now, let us try to reformulate the expansion law with invariant volume by preserving the fundamental nature of Planck's constant. Looking at the time derivative of invariant volume, which can be simplified to give
\begin{equation}
\frac{dV_k}{dt}=3HV_k-Ar_A^2\frac{\ddot{a}}{a},
\end{equation}
where $A=4\pi r_A^2$. Using the second Friedmann equation in Einstein gravity given by 
\begin{equation}\label{Freidmnn_2}
\frac{\ddot{a}}{a}=\frac{-4\pi}{3}(\rho+3P)l_p^2,
\end{equation}
one can express the difference in the number of degrees of freedom as 
\begin{equation}
\Delta N=\frac{A}{l_p^2}-\frac{3V_k}{l_p^2}\frac{\ddot{a}}{a}.
\end{equation}
With suitable rearrangements, the rate of change of invariant volume can be expressed as 
\begin{equation}\label{expasnion_new}
\frac{dV_k}{dt}=\frac{\bar{V}_k}{V_k}\big(l_p^2\Delta N\big)+3V_kH-\frac{A\bar{V}_k}{V_k}.
\end{equation}
This shows that the direct proportionality between $\frac{\Delta V}{\Delta t}$ and $\Delta N$ as envisioned in the original expansion law has been broken, when one employs the proper invariant volume instead of areal volume. A similar conclusion was made by the authors of \cite{modthanu_critic} that the rate of change of invariant volume is not proportional to $\Delta N$ unless one defines a complicated time-dependent Planck length. Thus the use of invariant volume to formulate the expansion law in non-flat universe  does not preserve the basic idea presented in Padmanabhan's conjecture. As noted earlier the first law of thermodynamics can be considered as the backbone of the expansion law. So the failure in formulating the expansion law using invariant volume suggests that the problem is much deeper, such that even the first law of thermodynamics may face similar problems when formulated using invariant volume.

\section{Feasibility of First Law of Thermodynamics in Non-Flat Universe}\label{2.1}
The authors of \cite{caiakb} have shown that the differential form of the Friedmann equation can be rewritten as a thermodynamic relation,
\begin{equation}\label{frst_law}
TdS=dE-WdV,
\end{equation} often designated as the unified first law of thermodynamics. Here E is the total energy of matter inside the apparent horizon $(E = \rho V)$, S is the entropy, $V$ is the volume of space inside the apparent horizon, W is the work density ($W=(\rho-p)/2$), and T is the temperature associated with the apparent horizon determined by the surface gravity at the horizon. However, Eq.\eqref{frst_law} was formulated using areal volume; therefore, one should check the status of the first law in the non-flat universe when one uses invariant volume.

Following Bakenstein's proposal, the entropy and the temperature of the horizon can be defined as
\begin{align}\label{Cai_T}
T&=\frac{-1}{2\pi r_A}(1-\frac{\dot{r_A}}{2Hr_A}),&
S&=\frac{A}{4l^2_p}
\end{align}
which gives $TdS$ as
\begin{equation}\label{lhs}
TdS = \frac{-dr_A}{l^2_p}\biggr(1-\frac{\dot{r_A}}{2Hr_A}\biggr).
\end{equation}   
This $TdS$ term serves as the L.H.S of Eq.\eqref{frst_law} and is independent of the choice of volume. But the R.H.S of Eq.\eqref{frst_law}  depends explicitly on the choice of volume.  Considering the invariant volume, its infinitesimal change as time progress is given by \cite{modthanu},
\begin{equation}\label{dv}
dV_k=(3V_kH-4\pi r_A^2)dt+\frac{4\pi r_A\dot{r_A}}{H}dr_A.
\end{equation}
Plugging in relation \eqref{dv} into Eq.\eqref{frst_law} leads us to the relation,
\begin{multline}\label{rhs1}
dE-WdV_k = V_kd\rho\\ +\frac{3}{2}(\rho+p)Hdt \biggr((V_k-\frac{4\pi r_A^2}{3H}+\frac{4\pi r_A \dot{r}_A}{3H^2}\biggr).
\end{multline}
From the continuity equation, we can use
\begin{equation}
d\rho= -3H(\rho+p)dt,
\end{equation}
to simplify the relation \eqref{rhs1} and yield
\begin{equation}\label{rhs2}
dE-WdV_k=\biggr(V_k+\frac{4\pi r_A^2}{3H}-\frac{4\pi r_A \dot{r}_A}{3H^2}\biggr)\frac{d\rho}{2}.
\end{equation}
Using the differential form of the first Friedmann equations given by
\begin{equation}\label{drho}
dr_A=-\frac{4\pi Gr_A^3}{3} d\rho,
\end{equation}
we can reduce Eq.\eqref{rhs2} to 
\begin{equation}\label{rhs3}
dE-WdV_k=-\frac{dr_A}{l_p^2}\biggr(\frac{V_k}{2\bar{V}_k}-\frac{1}{2Hr_A}\biggr)+\frac{TdS}{H r_A}.
\end{equation}
According to this equation, $dE-WdV_k,$ the R.H.S. of the unified first law as given by Eq.\eqref{frst_law} will not precisely
reduce to $TdS$, instead there appears an extra term proportional to
$\displaystyle \left(\frac{V_k}{2\bar{V}_k}-\frac{1}{2Hr_A}\right)$ which negate the conventional form of the first law. But, for flat universe, where $V_k =\bar{V}_k$ and $H=r_A^{-1},$ the additional term vanishes and  $dE-WdV_k$ will reduce to $TdS$ and thus guarantees the perfect validity of the first law. Thus the unified first law of thermodynamics with invariant volume can only be legitimate for a flat universe.

One may suspect that by treating the Planck length as a function of cosmic time as done in Ref.\cite{modthanu}, it may be possible to safeguard the first law of thermodynamics. By using such a time dependent Planck length (as defined using Eq.(10) in Ref.\cite{modthanu}) in Eq.\eqref{Cai_T}, the $TdS$ term will be modified as 
\begin{equation}\label{TdS_modfied}
TdS=\frac{-1}{2l^2_p}\biggr(1-\frac{\dot{r_A}}{2Hr_A}\biggr)\bigg(\frac{2f_kdr_A-r_Adf_k}{f_k^2}\bigg).
\end{equation}
Proceeding as previously the RHS of the first law takes the form
\begin{multline}\label{rhs_modifed}
dE-WdV_k=\frac{TdS}{H r_A}-\biggr(1-\frac{\dot{r_A}}{2Hr_A}\biggr)\frac{df_k}{l_p^2f_k^2H}\\-\frac{1}{2l_p^2}\biggr(\frac{V_k}{2\bar{V}_k}-\frac{1}{2Hr_A}\biggr)\bigg(\frac{2f_kdr_A+r_Adf_k}{f_k^2}\bigg)
\end{multline}
According to this equation, the RHS of the first law here also does not reduces to $TdS$, but there appears two additional terms. Similar to that of Eq.\eqref{rhs3}, one term is proportional to $\displaystyle \left(\frac{V_k}{2\bar{V}_k}-\frac{1}{2Hr_A}\right),$ and in addition to that a second term proportional to $df_k/l_p^2H$ is also present. For a flat universe, the additional terms will vanish since $f(t)_{k=0}=1$ and ${df_{k=0}}=0$, and the RHS of Eq.\eqref{rhs_modifed} will reduces to $TdS$.Thus, even though the time depended Planck length as proposed in \cite{modthanu} may safeguard the
expansion law with invariant volume to some extent, our results shows that it cannot safeguard the first law of thermodynamics. Thus, it is impossible to formulate the unified first law of thermodynamics with the use of invariant volume in non-flat universe.

\section{First Law of the Form $-dE=TdS$ at the Apparent Horizon Surface}\label{3}

Now we can turn our attention to another form of first law of thermodynamics proposed by the authors of \cite{Cai_first_law_friedmann_equations} and examine how the choice of volume will effect the formulation of first law. The heat flow  $\delta Q$ through the apparent horizon in an infinitesimal interval of time $dt$, is related to the change in the energy $-d\check{E}$ inside the apparent horizon by the relation $\delta Q=-d\check{E}$. It is important to point out that the energy change $d\check{E}$ considered here is quite different from the energy change $dE$ mentioned in the unified first law ($dE=TdS-WdV$). Here, $d\check{E}$ is the heat flux $\delta Q$  crossing the apparent horizon within an infinitesimal interval of time $dt$ \cite{Cai_first_law_friedmann_equations} and is given as 
\begin{equation}\label{frst_law_1}
d\check{E}=-A(\rho+p)r_AHdt.
\end{equation}
In calculating the energy flux, the size of the apparent horizon is assumed to be fixed. Similarly, the apparent horizon is assumed to have a temperature $\check{T}$ and entropy $S$ given by 
\cite{Cai_first_law_friedmann_equations}
\begin{align}\label{T1}
\check{T}&=\frac{1}{2\pi r_A},& S&=\frac{A}{4l_p^2}.
\end{align}
Therefore, from the energy-entropy relation, the first law of thermodynamics can be written as 
\begin{equation}\label{Frst_law1}
-d\check{E}=\check{T}dS.
\end{equation}

Considering this first law of thermodynamics as the backbone of the expansion law, the authors of \cite{expasnion_law_fom_frst_law} have derived the modified expansion law due to Sheykhi in (3+1) Einstein gravity. Here,    
we extend that approach to derive the expansion law in (n+1) Einstein gravity and further extend it to Gauss-Bonnet and more general Lovelock gravity.
In (n+1) dimensional Einstein gravity, the modified expansion law is given as  \cite{sheykhi}
\begin{equation}\label{expasnion_law_n-D}
\alpha\frac{dV}{dt}=l_p^{n-1}\frac{r_A}{H^{-1}}\big(N_{sur}-N_{bulk}\big),
\end{equation} 
where $\alpha=\frac{n-1}{2(n-2)}$ and $V=\Omega_nr_A^n$ is the areal volume of n-sphere. For (n+1) dimensional Einstein gravity, the $N_{sur}$ and $N_{bulk}$ are defined as 
\begin{align}\label{degrees_freedom_n-D}
N_{sur}&=\alpha\frac{A}{l_p^2},& N_{bulk}&=-4\pi\Omega_nr_A^{n+1}\frac{\big((n-2)\rho+np\big)}{n-2},
\end{align}
where, $A=n\Omega_nr_A^{n-1}$ is the surface area spanned by the horizon.
With the help of Eq.\eqref{frst_law_1}, the first law of thermodynamics for (n+1) Einstein gravity  can be represented mathematically  as 
\begin{equation}\label{frst_law_n}
\frac{1}{l_p^{n-1}}\frac{dr_A}{dt}=\frac{8\pi r_A^3(\rho+p)H}{n-1}.
\end{equation}
Utilising the n-dimensional Friedmann equation given by
\begin{equation}
H^2+\frac{k}{a^2}=\frac{16\pi l_p^{n-1}}{n(n-1)}\rho
\end{equation}
and Eq.\eqref{radius}, the relation \eqref{frst_law_n} can be modified to give, 
\begin{equation}\label{expasn_n1}
\frac{1}{l_p^{n-1}}\frac{dr_A}{dt}=r_AH\bigg(\frac{1}{l_p^{n-1}}+\frac{8\pi r_A^2(\rho+p)}{n-1}-\frac{16\pi r_A^2\rho}{n(n-1)}\bigg).
\end{equation} 
Multiplying both sides of the above equation with $n\Omega_nr_A^{n-1}$ and with some suitable rearrangements we get 
\begin{multline}\label{expnsion_nD_lststep}
\frac{n-1}{2(n-2)}\frac{dV}{dt}=l_p^{n-1}\frac{r_A}{H}\biggr(\frac{n-1}{2(n-2)}\frac{n\Omega_nr_A^{n-1}}{l_p^{n-1}}\\+4\pi\Omega_n r_A^{n+1}\frac{\big((n-2)\rho+np\big)}{n-2}\biggr).
\end{multline}
Employing the relations in \eqref{degrees_freedom_n-D}, one can express Eq.\eqref{expnsion_nD_lststep} as
\begin{equation}
\alpha\frac{dV}{dt}=l_p^{n-1}\frac{r_A}{H^{-1}}\big(N_{sur}-N_{bulk}\big)
\end{equation}
Which is nothing but the expansion law for (n+1) Einstein gravity.

Gauss-Bonnet gravity is a special higher-order gravity theory obtained by the modification of Einstein-Hilbert action by including the Gauss-Bonnet term given as  \cite{caiakb}
\begin{equation}
R_{GB}=R^2-4R^{\nu\mu}R_{\nu\mu}+4R^{\nu\mu\rho\sigma}R_{\nu\mu\rho\sigma}.
\end{equation}
The additional term presented here is a topological term and have only relevance in (4+1) dimensions or higher. It is a well known fact that the entropy relation \eqref{Cai_T} assumed in  Einstein gravity does not hold in higher-order gravity theories. Hence to account for the entropy associated with apparent horizon, one can assume the entropy relation for black hole horizon also holds for the apparent horizon of FRW universe and define the entropy in Gauss-Bonnet gravity as \cite{caiakb}
\begin{equation}\label{entropy_gauss_bonnet}
S=\frac{A}{4l_p^{n-1}}\bigg(1+\frac{n-1}{n-3}\frac{2\check{\alpha}}{r_A^2}\bigg),
\end{equation} 
where $\check{\alpha}=(n-2)(n-3)\alpha$ is known as the Gauss-Bonnet coefficient.
From Eq.\eqref{entropy_gauss_bonnet}, the effective area corresponding to the holographic surface can be defined as \cite{sheykhi}
\begin{equation}\label{effective_area}
\check{A}=A\bigg(1+\frac{n-1}{n-3}\frac{2\check{\alpha}}{r_A^2}\bigg),
\end{equation}
Now, we move on to deduce the expansion law from the first law of thermodynamics.
Employing the entropy relation \eqref{entropy_gauss_bonnet} and from Eq.\eqref{frst_law_1}, the first law of thermodynamics in Gauss-Bonnet gravity can be presented as  
\begin{equation}\label{frst_law_gauss_bonnet}
\frac{1}{l_p^{n-1}}\big(1+2\check{\alpha}r_A^{-2}\big)\frac{dr_A}{dt}=\frac{8\pi r_A^3(\rho+p)H}{n-1}.
\end{equation}
Using the Friedmann equation in Gauss-Bonnet gravity given by  \cite{Cai_first_law_friedmann_equations}
\begin{equation}\label{frdman_eqn_gauss}
H^2+\frac{k}{a^2}+\check{\alpha}\bigg(H^2+\frac{k}{a^2}\bigg)^2=\frac{16\pi l_p^{n-1}}{n(n-1)}\rho,
\end{equation}
and the relation \eqref{radius}, one can rewrite Eq.\eqref{frst_law_gauss_bonnet} as
\begin{multline}\label{expasnio_gauss1}
\frac{1}{l_p^{n-1}}\big(1+2\check{\alpha}r_A^{-2}\big)\frac{dr_A}{dt}=\\r_AH\bigg(\frac{1}{l_p^{n-1}}\big(1+2\check{\alpha}r_A^{-2}\big)+\frac{8\pi r_A^2(\rho+p)}{n-1}-\frac{16\pi l_p^{n-1}}{n(n-1)}\rho\bigg)
\end{multline}
Multiplying both sides with $n\Omega_nr_A^{n-1}$ and with some suitable rearrangements Eq.\eqref{expasnio_gauss1} reduces to 
\begin{multline}\label{expasnion Gauss2}
\frac{\alpha n\Omega_nr_A^{n-1}(1+2\check{\alpha}r_A^{-2})}{l_p^{n-1}}\frac{dr_A}{dt}=\\l_p^{n-1}\frac{r_A}{H^{-1}}\bigg(\alpha\frac{n\Omega_nr_A^{n-1}(1+2\check{\alpha}r_A^{-2})}{l_p^{n-1}}+\\4\pi\Omega_nr_A^{n+1}\frac{((n-2)\rho+np)}{n-2}\bigg)
\end{multline}
Let us define $N_{sur}$ for Gauss-Bonnet gravity as 
\begin{equation}
N_{sur}=\alpha\frac{n\Omega_nr_A^{n-1}(1+2\check{\alpha}r_A^{-2})}{l_p^{n-1}},
\end{equation}
which is identical to the relation for $N_{sur}$ defined in Ref. \cite{sheykhi}.
Similarly, from the relation \eqref{effective_area} we can define the effective increase in volume in Gauss-Bonnet gravity as\cite{sheykhi}
\begin{equation}
\frac{\check{dV}}{dt}=\frac{r_A}{n-1}\frac{d\check{A}}{dt}=n\Omega_nr_A^{n-1}(1+2\check{\alpha}r_A^{-2}).
\end{equation} 

Hence Eq.\eqref{expasnion Gauss2} can be conveniently represented as  
\begin{equation}
\alpha\frac{d\check{V}}{dt}=l_p^{n-1}\frac{r_A}{H^{-1}}\big(N_{sur}-N_{bulk}\big),
\end{equation}
which is the expansion law proposed in \cite{sheykhi} to derive the Friedmann equations in Gauss-Bonnet gravity.

Now we consider the more general Lovelock gravity theory, which is the generalisation of Einstein gravity when space-time assumes a dimension greater than four. Assuming the entropy relation for black-holes in Lovelock gravity holds for the apparent horizon of the FRW universe, the entropy associated with the apparent horizon can be defined as \cite{Cai_first_law_friedmann_equations} 
\begin{equation}\label{entropy_lovlck}
S=\frac{A}{l_p^{n-1}}\Sigma_{i=1}^{m}\frac{i(n-1)}{n-2i+1}\hat{c_i}r_A^{2-2i}.
\end{equation}
where $m=n/2$ and the coefficients are given as
\begin{equation}
\begin{split}
\hat{c_0}=\frac{c_0}{n(n-1)}, \text{\quad} \hat{c_1}=1,\\ \hat{c_i}=c_i\Pi^{2m}_{j=3}(n+1-j),  \text{\quad for\quad } i>1.
\end{split}
\end{equation} 
By defining the effective area like in the previous case, the rate of change of effective volume in Lovelock gravity can be found as \cite{sheykhi}
\begin{equation}\label{effective_vol_lvlck}
\frac{\check{dV}}{dt}=\frac{r_A}{n-1}\frac{d\check{A}}{dt}=n\Omega_nr_A^{n+1}\Sigma_{i=1}^{m}\bigg(i\hat{c_i}r_A^{-2i}\bigg).
\end{equation}
Moving onto our plan, the first law of thermodynamics in Lovelock gravity can be expressed with the help of Eq.\eqref{frst_law_1} and Eq.\eqref{entropy_lovlck} as
\begin{equation}\label{frst_law_lvlck1}
\frac{1}{l_p^{n-1}}\frac{dr_A}{dt}\Sigma_{i=1}^{m}\bigg(i\hat{c_i}r_A^{2-2i}\bigg)=\frac{8\pi r_A^3(\rho+p)H}{n-1}.
\end{equation}
Using the Eq.\eqref{radius} and the Friedmann equation in Lovelock gravity given by \cite{Cai_first_law_friedmann_equations}
\begin{equation}\label{frdman_eqn_lvlck}
\Sigma_{i=1}^{m}\hat{c_i}\bigg(H^2+\frac{k}{a^2}\bigg)^i=\frac{16\pi l_p^{n-1}}{n(n-1)}\rho,
\end{equation}
one can rewrite Eq.\eqref{frdman_eqn_lvlck} as
\begin{multline}\label{lvlck_frst_law}
\frac{1}{l_p^{n-1}}\frac{dr_A}{dt}\Sigma_{i=1}^{m}\bigg(i\hat{c_i}r_A^{2-2i}\bigg)=r_AH\bigg(\frac{\Sigma_{i=1}^{m}\big(\hat{c_i}r_A^{2-2i}\big)}{l_p^{n-1}}\\+\frac{8\pi r_A^3(\rho+p)H}{n-1}-\frac{16\pi l_p^{n-1}}{n(n-1)}\rho\bigg).
\end{multline}
Multiplying both sides with $n\Omega_nr_A^{n+1}$ and with some suitable rearrangements Eq.\eqref{lvlck_frst_law} can be presented as 
\begin{multline}
\alpha\frac{n\Omega_nr_A^{n+1}}{l_p^{n-1}}\frac{dr_A}{dt}\Sigma_{i=1}^{m}\bigg(i\hat{c_i}r_A^{2-2i}\bigg)=\\ \frac{1}{l_p^{n-1}}\frac{r_A}{H^{-1}}\bigg(n\alpha\Omega_nr_A^{n-1}\frac{\Sigma_{i=1}^{m}\big(\hat{c_i}r_A^{-2i}\big)}{l_p^{n-1}}\\+4\pi\Omega_nr_A^{n+1}\frac{((n-2)\rho+np)}{n-2}\bigg)
\end{multline}
Let us define $N_{sur}$ for Lovelock gravity as
\begin{align}\label{lvlck2}
N_{sur}&=\frac{1}{l_p^{n-1}}\frac{r_A}{H^{-1}}\bigg(n\alpha\Omega_nr_A^{n-1}\frac{\Sigma_{i=1}^{m}\big(\hat{c_i}r_A^{-2i}\big)}{l_p^{n-1}}\bigg),
\end{align}
which is identical to the relation for $N_{sur}$ in Lovelock gravity as proposed in Ref. \cite{sheykhi}. Thus with the use of relation \eqref{effective_vol_lvlck} one can conveniently write Eq.\eqref{lvlck2} as 
\begin{equation}
\alpha\frac{d\check{V}}{dt}=l_p^{n-1}\frac{r_A}{H^{-1}}\big(N_{sur}-N_{bulk}\big),
\end{equation}
which is the expansion law proposed in \cite{sheykhi} for the Lovelock gravity. Thus the above presented results strengthen the thermodynamic backbone of the expansion law formulated with areal volume.

In all the above derivations, we have considered the areal volume of the horizon for the non-flat universe to derive the expansion law following the method of Sheykhi. One can now think about the possibility of deriving an expansion from the first law of the form $-dE=TdS$ using the invariant volume of the horizon instead of the areal volume. However, before that, the deeper question addressed here is the status of first law itself when one uses the invariant volume.  In this context, it is important to have a closer look at the energy transfer across the apparent horizon within an infinitesimal interval of time $dt$. Let $V$ be the volume of the space inside the apparent horizon, and $d\rho$ is the  change in energy density in an infinitesimal interval of time $dt$, then corresponding total change in energy inside the volume is 
\begin{equation}
d\check{E_v}=Vd\rho
\end{equation}
In calculating the energy change within the volume, the size of the apparent horizon is assumed to be fixed. Using the continuity equation the change in energy within an infinitesimal interval of time $dt$ inside  the apparent horizon can be defined as,  
\begin{equation}\label{energy_flux}
d\check{E_v}=-3V(\rho+p)Hdt.
\end{equation} 
If $V$ is the areal volume of the horizon, then $d\check{E_v}$ will be identical with $d\check{E}$, the energy flux crossing the apparent horizon in an infinitesimal interval of time. On the other hand, if V is the invariant volume of the horizon, then $d\check{E_v}$  will be intrinsically different from $d\check{E}$. Thus, the energy flux across the surface  $d\check{E}$ and the energy change $d\check{E_v}$ within the volume enclosed by the surface are equivalent only when one uses areal volume for the horizon.

In Ref. \cite{caiakb}, it was argued that the unified first law ($dE=TdS+WdV$) and the first law of the form $dE=-TdS$ are consistent with each other even though the temperature and the definition of energy are different in each case. However, they agree with each other only if we use the areal volume.  The unified first law $dE = TdS + WdV$ cannot be formulated properly with invariant volume and the description of the first law of the form $dE = -TdS$ with the use of invariant volume cannot account for the energy transfer across the surface correctly. Hence, a consistent description of the first law of thermodynamics demands the use of areal volume. 

\section{Summary and Discussion}\label{5}
In this paper, we addressed the problems faced by choice of invariant volume to formulate an expansion law in the non-flat universe. The authors of \cite{modthanu} have formulated an expansion law with invariant volume for the non-flat universe, but they had to redefine the Planck length as a function of cosmic time to derive the correct Friedmann equations. Altering a fundamental constant is itself questionable, and moreover, the altered Planck length was divergent for radiation and the matter-dominated era of a non-flat open universe. Otherwise, this implies that, as we have shown, the emergence of cosmic space will not be proportional to the difference in degrees of freedom ($N_{sur}- N_{bulk}$) as envisaged by Padmanabhan. Our previous results in \cite{mahith} have shown that the first law of thermodynamics is the backbone of the expansion law. Thus the problems faced by the expansion law when one uses invariant volume suggests that the problem is further deeper, and one should suspect the feasibility of the first law itself while using invariant volume. Our investigation shows that it is impossible to formulate the unified first law of thermodynamics ($dE = TdS + WdV$) with the use of invariant volume for a non-flat universe. Moreover, even with the time-dependent Planck length, one cannot safeguard the unified first law of thermodynamics.

We then analyzed another well-known form of the first law of thermodynamics ($-d\check{E}    = \check{T}dS$) proposed by the authors of \cite{Cai_first_law_friedmann_equations} to see how the choice of volume will affect the feasibility of first law. Expansion law from the first law of the form    $-d\check{E}= \check{T}dS$ has been obtained for the non-flat universe with areal volume by the authors of \cite{expasnion_law_fom_frst_law}. We have extended this result to  (n+1) Einstein, Gauss-Bonnet and more general Lovelock gravity with the areal volume. This confirms the deep connection between the law of emergence and the first law of thermodynamics in the form $-d\check{E}=\check{T}dS$. Hence, the difficulties faced in formulating the expansion law must be reflected in the formulation of first law (,$-d\check{E}=\check{T}dS$) using proper invariant volume, even though this first law does not explicitly seem to depend upon the volume term. Our analysis shows that the energy flux crossing the apparent horizon is equal to the change in energy within the volume bounded by the horizon, only if one uses the areal volume. We have further shown that the claimed consistency between the unified first law of the form $dE=TdS+WdV$ and $-d\check{E}=\check{T}dS$ in Ref. \cite{caiakb} will hold only with the use of areal volume. Thus the feasibility of the first law of thermodynamics in a non-flat universe necessitates the use of areal volume.  On the other hand, it may also be indicating that the universe may be flat as advocated by the current observations.

\bibliographystyle{apsrev4-1}
\end{document}